\documentclass[aps,twocolumn,prb,showpacs,eqsecnum]{revtex4-1}
\usepackage{graphpap}
\usepackage[dvips]{graphicx}
\usepackage[dvips]{graphics}
\usepackage{color}
\usepackage{multirow}

\newcommand{\specialcell}[2][c]{%
  \begin{tabular}[#1]{@{}c@{}}#2\end{tabular}}

\begin{document}

\title{Dielectric screening of the Kohn anomaly of graphene on hexagonal boron nitride}
\author{F. Forster$^{1,2}$, A. Molina-Sanchez$^{3,4}$, S. Engels$^{1,2}$, A. Epping$^{1,2}$,  K. Watanabe$^5$, T. Taniguchi$^5$, L. Wirtz$^{3,4}$, and C. Stampfer$^{1,2}$}

 \affiliation{
$^1$JARA-FIT and II. Institute of Physics, RWTH Aachen University, 52074 Aachen, Germany, EU \\
$^2$Peter Gr\"unberg Institute (PGI-9), Forschungszentrum J\"ulich, 52425 J\"ulich, Germany, EU \\
$^3$Institute for Electronics, Microelectronics, and Nanotechnology (IEMN), CNRS UMR 8520, Dept. ISEN, 59652 Villeneuve d'Ascq Cedex, France, EU \\
$^4$Physics and Materials Science Research Unit, University of Luxembourg, L-1511 Luxembourg, Luxembourg, EU \\
$^5$National Institute for Materials Science, 1-1 Namiki, Tsukuba, 305-0044, Japan
}

\begin{abstract}

Kohn anomalies in three-dimensional metallic crystals are dips in the phonon dispersion that are caused by abrupt changes in the screening of the ion-cores by the surrounding electron-gas. These anomalies are also present at the high-symmetry points $\Gamma$ and K in the phonon dispersion of two-dimensional graphene, where the phonon wave-vector connects two points on the Fermi surface.  The linear slope around the kinks in the highest optical branch is proportional to the electron-phonon coupling. Here, we present a combined theoretical and experimental study of the influence of the dielectric substrate on the vibrational properties of graphene. We show that screening by the dielectric substrate reduces the electron-phonon coupling at the high-symmetry point K and leads to an up-shift of the Raman 2D-line. This results in the observation of a Kohn anomaly that can be tuned by screening. The exact position of the 2D-line can thus be taken also as a signature for changes in the (electron-phonon limited) conductivity of graphene.

\end{abstract}


\date{ \today}

\maketitle

\section{Introduction}
Graphene, 
 a monoatomic carbon
membrane with unique electronic properties~\cite{kat12,cas09} is a promising candidate for flexible electronics,
high frequency applications and spintronics~\cite{coo12}. 
However, graphene's ultimate surface-to-volume ratio makes the environment,
in particular the substrate material, have a pronounced influence
onto its intrinsic properties.
For example, 
SiO$_2$, the most common substrate material,
exhibits surface roughness, dangling bonds and charge traps which 
  introduce ripples, disorder~\cite{mar07}, and doping domain fluctuations~\cite{sta07}. This limits carrier 
mobilities and the operation of graphene devices~\cite{pon09,wan11}. 
Therefore alternative substrates 
are required to overcome these limitations. Hexagonal boron nitride (hBN) has been identified as
a very promising candidate~\cite{dea10, may11, xue11, dec11}. A large (indirect) band gap,
a lattice mismatch to graphite of less than 2\%, and the
absence of dangling bonds makes this atomically
flat material a valuable and promising insulating counter part to graphene~\cite{arnaud06,latconst}.
It has been shown that graphene can be successfully 
transferred to ultra-thin hBN flakes leading to improved electronic
transport properties compared to graphene on SiO$_2$~\cite{dea10, may11}. 
Moreover, scanning tunneling microscopy experiments have shown
that the sizes of individual electron-hole puddles are 
significantly increased while the disorder potential 
is reduced by roughly a factor ten~\cite{xue11,dec11}. 

Over the last years
Raman spectroscopy has proven to be a powerful tool 
for characterizing graphene and studying its physical properties. 
For example, this technique has been successfully used (i) to
distinguish single-layer graphene from few-layer graphene
and graphite~\cite{fer06,dav07a,gupta06},
(ii) to monitor doping levels~\cite{pis07,sta07}, (iii) to study short range disorder and edge properties~\cite{cinziaNL}
and (iv) to investigate suspended~\cite{berciaud} and nanostructured graphene~\cite{bis11}.
Very recently, Raman measurements of graphene deposited on a hexagonal boron nitride (hBN) substrate have displayed subtle changes with respect to graphene on SiO$_2$~\cite{wang12,hone12}. The G-line was shown to down-shift slightly by about 4 cm$^{-1}$, a behaviour that has also been observed for suspended graphene~\cite{berciaud}. This red-shift was explained by the reduced doping level of 
suspended graphene and graphene on the pure hBN substrate. The 2D-line, however
displays opposite behaviour in the two cases, it displays a red-shift in 
suspended graphene and a blue-shift for graphene on hBN (and even more so for graphene embedded in hBN~\cite{hone12}). 
These shifts in opposite
directions cannot be explained by the absence of impurities and have remained
a puzzle up to now. The resolution of this effect is the object of the
current article. We show here through a combined experimental and theoretical
approach that the monoatomic layered structure of graphene renders the Kohn 
anomaly~\cite{Kohn59,piscanec} at the high-symmetry point $K$ susceptible to the screening by the 
dielectric substrate. We present spatially resolved confocal Raman 
spectroscopy measurements of graphene on hexagonal boron nitride substrates
which are compared with measurements of graphene on SiO$_2$.

\begin{figure}[t]\centering
\includegraphics[keepaspectratio=true,clip,%
                   width=0.7\linewidth]%
                   {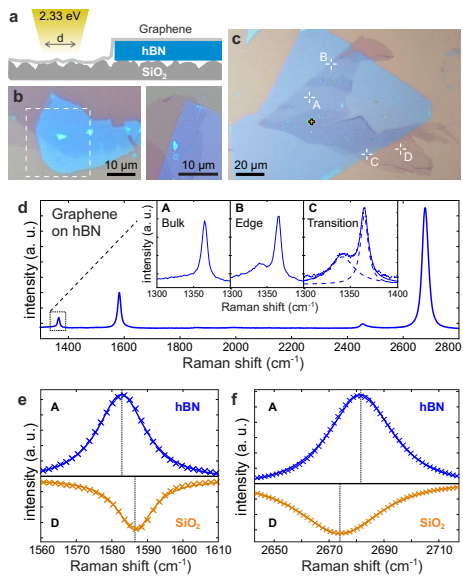}                   
\caption[FIG1]{
({a}) Schematic illustration of a graphene sample with incident laser. ({b-c}) Optical microscope images of graphene flakes partly resting on hBN (blue) and SiO$_2$. The flake consists of different regions of single-layer, bilayer and few-layer graphene. ({d}) Raman spectrum of single-layer graphene resting on hBN. Inset: Region around the hBN peak at sites marked in panel c. ({e-f}) G ({e}) and 2D ({f}) peak of single-layer graphene on hBN (marker A) and SiO$_2$ (marker D).} 
\label{samples}
\end{figure}

\section{Measurements and results}
\subsection{Sample fabrication}

Graphene and hBN flakes are prepared by micromechanical cleavage. 
While hBN is directly deposited on a SiO$_2$/Si$^{++}$ substrate, graphene is prepared on top of a polymer stack consisting of a water-soluble polymer [100~nm Polyvinylalcohol (PVA)] and a water resistant polymer [270~nm Polymethyl methacrylate (PMMA)] allowing the transfer process described in detail in Ref.~8. 
 Before depositing the graphitic flakes on top of the hBN, Raman spectroscopy has been used to identify and select individual single-layer graphene flakes~\cite{fer06,dav07a}. 
The Raman data are recorded by using a laser excitation of 532~nm ($E_L$=2.33~eV) through a single-mode optical fiber whose spot size is limited by diffraction. A long working distance focusing lens with numerical aperture of 0.85 is used to obtain a spot size of approximately\ 400~nm. 
We used a laser power below 1~mW such that heating effects can be 
neglected~\cite{jun07a}.

\subsection{Raman spectroscopy measurements}
A schematic illustration of our structures is shown in Fig.~1a and optical microscope images of some fabricated samples are shown in Figs.~1b-c.
A typical Raman spectrum of a graphene flake resting on hBN is shown in Fig.~1d. The Raman spectrum shows the prominent G-line around 1582~cm$^{-1}$ as well as the single Lorentzian shaped 2D-line around 2675~cm$^{-1}$ as expected for graphene. A third prominent and sharp peak arises around 1365~cm$^{-1}$, which can be attributed to 
the Raman active LO-phonon in hBN~\cite{gei66}.
It is important to distinguish this peak from 
the defect induced D-line potentially appearing at around 1345~cm$^{-1}$~\cite{dav07a}.
Therefore, Raman spectra have been acquired at edge regions of the graphene flake where defects are known to appear \cite{dav07a}. The insets in Fig.~1d show corresponding Raman spectra at different positions marked and labeled in Fig.~1c. The data recorded at the edge (B) shows a second peak arising at around 1345~cm$^{-1}$ which is not visible in the bulk region and can be clearly distinguished from the one at 1365~cm$^{-1}$. 
 As shown in inset C of Fig.~1d, the D-line also appears in regions of substrate transition, i.e.~ at the edge of the underlying hBN flake. This can be attributed to local bending of the graphene flake induced by the level difference of the two substrates. The sp$^{2+\eta}$ hybrid orbitals necessary to bend the graphene layer cause short range scattering leading to a D peak in the Raman spectrum. The D line is not visible in regions of graphene away from the edges,
neither on hBN nor on SiO$_2$. We conclude that the transfer technique used in the fabrication process does not induce a significant amount of defects in the graphene lattice. 

In order to show the substrate dependence of the Raman lines of graphene, 
we compare in Figs.~1e-f  
the G and 2D-lines of a flake that partially rests on hBN and partially
on SiO$_2$. The G peak of graphene on hBN and the one on SiO$_2$ 
differ significantly in their 
position: the G peak on SiO$_2$ is centered at 1586.5~cm$^{-1}$ while
the one on hBN is centered at 1582.8~cm$^{-1}$. This downshift can be
attributed to reduced doping, which also is consistent with the increase
of the full width at half maximum (FWHM) of the G peak of graphene on hBN~\cite{yan07}.
The FWHM is 12.2~cm$^{-1}$ on SiO$_2$ and 16.7~cm$^{-1}$ on hBN. Also the 2D peak shows a 
substrate dependence of the position which is 2674.0~cm$^{-1}$ on SiO$_2$ 
and 2681.6 ~cm$^{-1}$ on hBN. The substrate dependence of the 2D FWHM is even 
more significant, being 36.4~cm$^{-1}$ on SiO$_2$ and 28.1~cm$^{-1}$ on hBN.

\begin{figure}[t]\centering
\includegraphics[keepaspectratio=true,clip,%
                   width=0.65\linewidth]%
                   {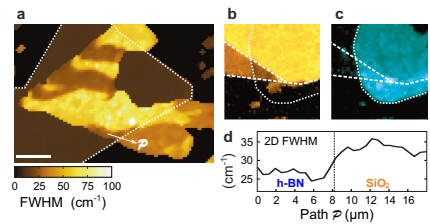}                   
\caption[FIG2]{
({a}) 2D FWHM Raman image of the sample shown in Fig.~1c. Where no 2D peak was found the hBN peak FWHM was plotted making the hBN flake visible (highlighted by white dotted lines). ({ b}) Raman map of the 2D FWHM of the sample shown in Fig.~1b (see white dashed box in Figure~1b). Dotted line marks the hBN substrate. ({c}) Raman map of the same sample but integrated intensity of the hBN peak. ({d}) Line cut along the solid arrow (path P) in panel a.}
\label{2Dmaps}
\end{figure}

The peak-shifts and changes in the FWHM can not only be seen in individual 
Raman spectra, but also appear spatially resolved in two dimensional Raman 
maps. A Raman map of the 2D FWHM of the sample presented in Fig.~1c 
is shown in Fig.~2a. One can identify three single layer regions with a FWHM below 40~cm$^{-1}$, two resting on hBN and one resting on SiO$_2$ with a small region also resting on hBN. A line cut in this substrate transition region shown in Fig.~2d reveals a locally resolved difference of the FWHM of around 8~cm$^{-1}$. Fig.~2b shows the 2D FWHM map of the sample previously shown in Fig.~1b (left panel). There is also a substrate dependency visible. 
 A Raman map of the integrated peak intensity between 1360 and 1370~cm$^ {-1}$ is shown in Fig.~2c. The bright area has exactly the same shape as the underlying hBN flake in the optical picture and verifies the attribution of this peak to the hBN mode. 

\begin{figure}[t]\centering
\includegraphics[keepaspectratio=true,clip,%
                   width=0.65\linewidth]%
                   {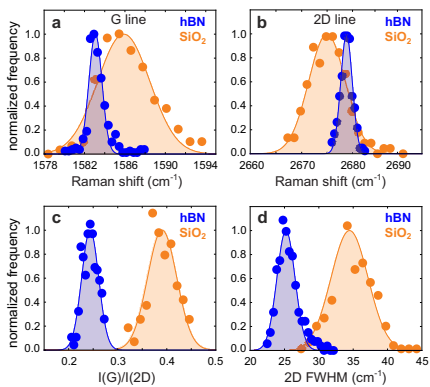}                   
\caption[FIG1]{
Statistical evaluation of single-layer regions of the Raman map shown in Fig.~2a in terms of G peak position ({a}), 2D peak position ({b}), intensity ratio between G and 2D peak ({c}), and FWHM of the 2D peak ({d}).
}
\label{statistics}
\end{figure}

\subsection{Statistical analysis}
To relate the quantitative results of the measurements to physical properties 
like charge carrier density fluctuations and to dispose of statistical 
fluctuations, it is necessary to evaluate a large number of spectra 
statistically.
The statistical distribution of the G peak position of the flake of 
Fig.~1c, only considering single-layer regions, is plotted in 
Fig.~3a. These data show a clear distinction between the different substrates. To obtain estimates of the statistical parameters the distribution is considered to be approximately Gaussian. The mean value of the distribution on hBN, $\mu_\mathrm{hBN}$=1583.1~cm$^{-1}$, is red shifted with respect to the one on SiO$_2$, $\mu_\mathrm{SiO_2}$=1585.9~cm$^{-1}$, a notable deviation by almost three wavenumbers. Even more pronounced is the difference in the
 standard deviation which is a measure for the G peak fluctuations. Being  $\sigma_\mathrm{hBN}$=0.7~cm$^{-1}$ on hBN, it is almost four times smaller than on SiO$_2$ where $\sigma_\mathrm{SiO_2}$=2.7~cm$^{-1}$.

Previous Raman measurements with gated graphene flakes on SiO$_2$ have demonstrated a dependency of the G peak position on the charge carrier density~\cite{das08,sta07}. A non-adiabatic theory was established to calculate the G-peak shift in terms of charge carrier densities~\cite{laz06,pis07}. By using a finite temperature of 295~K and an intrinsic G peak position of 1582.5~cm$^{-1}$ according to~\cite{das08}, we obtain a charge carrier density of 1.8$\times 10^{12}$~cm$^{-2}$ on SiO$_2$ and 9$\times 10^{11}$~cm$^{-2}$ on hBN, meaning the overall doping of the investigated single-layer flake on hBN is reduced by a factor of two with respect to a region of the very same flake resting on SiO$_2$.
Please note that the doping induced G-peak shifts are well consistent with the observed FWHMs of the corresponding G-lines~\cite{sta07,yan07}.

Furthermore, using the standard deviation of the G-peak-shift distribution to quantify the charge fluctuations, one obtains a fluctuation of 1.6$\times 10^{12}$~cm$^{-2}$ on SiO$_2$ and a fluctuation of 6$\times 10^{11}$~cm$^{-2}$ on hBN. This difference by almost a factor of three indicates a significant reduction in doping domain fluctuations and hence a reduction of the disorder potential in the single-layer graphene on hBN. Comparing these results with the data obtained by scanning tunneling spectroscopy~\cite{xue11,dec11}, one will notice a difference by one to two orders of magnitude. However, the experiments using scanning tunneling spectroscopy, besides being carried out in a low temperature controlled environment, acquire data with an atomic resolution from an area of 100~nm$^2$ which is on the scale of individual charge puddles, the Raman setup with a spatial resolution of around 500~nm is capable of measuring on a micrometer scale averaging over a large number of charge puddles. For instance, the areas used to acquire the data on this sample are around 50~$\mu$m$^2$ on SiO$_2$ and around 100~$\mu$m$^2$ on hBN.

As shown in previous works, the reduced doping fluctuations are also manifested in a reduced ratio between the integrated peak intensities of the G and 2D peak~\cite{das08,bas09}. The statistical distribution depicted in Fig.~3c shows a clear substrate dependency of this ratio and is in qualitative agreement with the G peak evaluation. 

Experimental data of the statistics of the FWHM of the 2D-line is provided in Fig.~3d. The mean values for the two substrates are clearly different and also the standard deviations differ. Consistent with the two dimensional Raman map shown before, the 2D peak width of the regions on hBN ($\mu_\mathrm{hBN}$=25.2~cm$^{-1}$) is significantly smaller by almost 10~cm$^{-1}$ than the one on SiO$_2$ ($\mu_\mathrm{SiO_2}$=34.5~cm$^{-1}$). Also the fluctuations of the hBN data, $\sigma_\mathrm{hBN}$=1.3~cm$^{-1}$, are significantly smaller than the ones of the SiO$_2$ data, $\sigma_\mathrm{SiO_2}$=2.7~cm$^{-1}$. So far, the FWHM of the 2D peak was not considered to be doping dependent. Measurements of gated graphene on SiO$_2$ showed no significant dependence on the charge carrier density~\cite{sta07}. Hence, a reduced doping alone cannot explain this substrate dependence. A similar reduced line-width of the 
2D-line (23 cm$^{-1}$) was observed for suspended graphene~\cite{berciaud,berciaud13}. We assume that the increased FWHM 
for graphite on SiO$_2$ is due to the substrate roughness and the presence of 
impurities which gives rise to an enhanced electron scattering and thus a smaller life-time of the excited electronic states during the double-resonant Raman process.\cite{discuss}

\section{Discussion}
We now turn to the discussion of the 2D-line position whose statistical distribution on hBN and on SiO$_2$ is shown in Fig.~3b.
While the position of the G-line can be directly related to the presence or absence of residual charging due to impurities on or in the substrate, the interpretation of the 2D-line shift is more subtle. Small charge densities ($<$ 4$\times$10$^{12}$ cm$^{-2}$) lead to shifts of the 2D-line by at most 2 cm$^{-1}$ (Ref.~5). 
 Furthermore, the 2D-line positions of suspended graphene~\cite{berciaud} and graphene on hBN differ by almost 10 cm$^{-1}$ even though both systems are mostly free of charge impurities, as demonstrated by the coincidence of the G-line positions. Thus, we discard charging as the origin for the 2D-line shift.

Our interpretation of the 2D-line shift is based on the double-resonance
Raman model of Thomsen and Reich~\cite{thomsen00}. The model successfully 
describes the D and 2D dispersion as a function of laser
energy as well as the splitting of the 2D-line for bilayer, and
few-layer graphene~\cite{fer06,dav07a,venezuela}, provided that 
renormalization of the highest optical phonon branch (HOB) due to 
electron-correlation effects is properly taken into account~\cite{lazzeri08}.
\begin{figure}[t]\centering
\includegraphics[keepaspectratio=true,clip,%
                   width=0.72\linewidth]%
                   {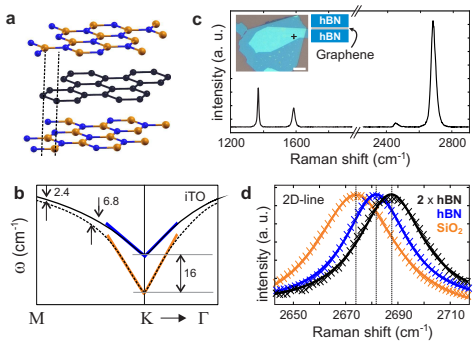}                   
\caption[FIG4]{
({a}) Geometry employed for the calculation of
the electron-phonon coupling in graphene surrounded by hBN. 
({b}) Schematic
figure of the differences in the highest-optical phonon branch around K
between suspended graphene and graphene surrounded by hBN.
({c}) Raman spectrum of graphene surrounded on 
both sides by multi-layer hBN. The sample is shown in the left inset.
The darker-blue area is the region of graphene on an underlying hBN 
flake. In the light-blue area, an additional hBN layer is deposited on 
top of the graphene layers. The spot where the spectrum was measured is 
marked by the black cross and the scale bar is 5~$\mu$m. ({d}) Raman
2D-line of graphene on SiO$_2$, graphene on hBN, and graphene 
surrounded by hBN.}
\label{double_and_calcfig}
\end{figure}
According to the double-resonance model, the 2D-line dispersion is 
proportional to the slope
of the HOB between the high-symmetry points K and M, and inversely proportional to the slope of the $\pi$ bands around K. In a first step, we thus calculated
the electronic band structure and the phonon-dispersion of graphene on 
hBN using standard density-functional theory in the local density approximation (DFT-LDA). We chose the most stable configuration where
one carbon atom is on top of a boron atom and the other carbon atom in the 
hollow site of hBN~\cite{giovannetti}. In the electronic band-structure
a small gap of 53 meV opens at K but further away from K,
the $\pi$-bands remain almost unchanged~\cite{giovannetti}. In the phonon dispersion, calculated
by density-functional perturbation theory (DFPT)~\cite{gonze}, there is only
a small change of the HOB in the immediate neighborhood around K. There, one 
observes a slight smearing of the Kohn anomaly (due to the very small 
band-gap opening), manifest in an up-shift of the HOB by about 3~cm$^{-1}$. 
Everywhere else between K and M, in particular in the wave-vector range that is 
sampled in Raman experiments, the upshift of the HOB is less than 1~cm$^{-1}$. 
We conclude that the pure ``mechanical'' interaction alone between graphene and 
the hBN substrate cannot explain the blue-shift of the Raman 2D-line~\cite{random_note}.

In recent work, it was shown by calculations on the level of the
GW-approximation~\cite{gwreview} that electronic correlation beyond DFT-LDA
influences both the slope of the $\pi$-bands of graphene~\cite{gruneis,trevisanutto} and the slope of the highest-optical phonon branch around K~\cite{lazzeri08}. The electron-electron interaction depends on the electronic screening
by the environment. Therefore, we expect that correlation effects in graphene
will be reduced by a dielectric substrate.
Although SiO$_2$ and hBN have both
roughly the same dielectric constant, the coupling between graphene and 
ultra-flat hBN is significantly increased compared to graphene on rough 
SiO$_2$ (see also illustration in Fig.~1a). This different dielectric
environment will have consequences
for the Fermi velocity and for the electron-phonon coupling.
In order to verify this hypothesis, we have performed GW-calculations
on isolated (suspended) graphene and on graphene surrounded by two layers of hBN.
The periodic geometry that we used in our calculations is shown in 
Fig.~4a. For simplicity, we have symmetrized the 
unit cell by choosing an ABC stacking sequence for the three layers. 
This ensures that 
the two carbon atoms in the unit-cell are equivalent, each with a boron atom 
on one side and a hollow site on the other side. Due to the symmetry, the 
linear crossing of the $\pi$-bands is preserved and we can use the same
strategy as in Ref.~31 
 for the calculation of the
electron-phonon coupling at the high-symmetry point K: In the $\sqrt{3}\times\sqrt{3}$ supercell, the atoms are displaced by a small distance ($d=0.01$~atomic units) 
from their equilibrium position along the eigenvector of the 
HOB (see Fig.~3b of Ref.~31). 
 The squared electron-phonon (e-ph) coupling -- which determines the slope
of the HOB around K~\cite{piscanec} -- is then obtained as
\begin{equation}
\left\langle D^2_{\mbox{\bf K}}\right\rangle = \frac{1}{8}
\left( \frac{\Delta E_{\mbox{\bf K}}}{d} \right)^2,
\end{equation}
where $\Delta E_{\mbox{\bf K}}$ is the induced energy gap between the $\pi^*$
and $\pi$ bands at K.

\begin{table}[t]
\begin{tabular}{l | c | c | c}
\hline\hline

				\multicolumn{2}{c|}{} 	 & \specialcell[t]{isolated  \\ graphene}      &   \specialcell[t]{graphene  \\ surrounded by hBN}        \\
\hline

\multirow{2}{*}{$\Delta E_{\mbox{\bf K}}$ (eV)}    &  LDA    & 0.1414     &  0.1388    \\
        		& GW & 0.2158      &  0.2070    \\
\hline
\multirow{2}{*}{$\left\langle D^2_{\mbox{\bf K}}\right\rangle$ (eV$^2$/\AA$^2$)} & LDA      & 89.25     &  86.00   \\
           & GW  &  207.88  &  191.27   \\
\hline
\end{tabular}
\caption{Calculated band-gap opening (for a displacement $d=0.0053$ {\AA}) and 
electron-phonon coupling of the highest optical ($A'_1$) phonon at K. 
Comparison of LDA and GW calculations.}
\label{output}
\end{table}

In Table~1, we present the results of our 
calculations (see Appendix for details). On the LDA level, the e-ph coupling
is 3.7\% weaker for graphene surrounded by hBN than for pure (suspended)
graphene. This difference
is increased to 8\% on the level of the GW approximation.
Obviously, the increased screening by the hBN substrate reduces the
gap opening and thus the e-ph coupling~\cite{note_gap_opening}. 
Since the slope of the HOB around
K is proportional to the e-ph coupling~\cite{piscanec} and since the
frequency of the HOB far away from K is almost independent of screening 
effects~\cite{lazzeri08}, a reduction
of the e-ph coupling leads to an increase of the HOB frequency at
and around K. This is demonstrated in Fig.~4b.
In order to quantitatively understand this effect,
it would be desirable to calculate the exact phonon dispersion relation
including correlation effects. Due to the complexity of total-energy 
GW calculations, this is currently not feasible. A workaround
consists in the use of the hybrid B3LYP functional~\cite{b3lyp} where
we adjust the parameters in order to mimic the two different screening
values for ``pure'' and ``sandwiched'' graphene (see details in the Appendix).
Using these functionals, we obtain that the HOB shifts 
(from ``pure'' to ``sandwiched'' graphene) by +16 cm$^{-1}$ at K, 
by +2.4~cm$^{-1}$ at M,
and by +6.8~cm$^{-1}$ half-way between K and M (which corresponds to a
excitation energy of 2.8~eV). At the same time, the frequency at 
$\Gamma$ remains almost unchanged (-0.4~cm$^{-1}$). For the laser energy
of 2.33~eV, linear interpolation yields a phonon shift of +8.3~cm$^{-1}$.
The corresponding 2D-line shift (where 2 phonons are excited/absorbed)
would be then +16.6~cm$^{-1}$. These calculations are not meant to provide
absolute numbers for the 2D-line shift but they demonstrate that the
experimentally measured 2D-line shift qualitatively agrees with the shift that
is induced by the enhanced dielectric screening for graphene on hBN~\cite{note_GWbands}.

So far, we have compared a calculation for graphene surrounded by hBN with
experiments where graphene is deposited on top of a hBN flake. 
This motivated us to perform Raman measurements on graphene surrounded
by hBN on both sides. We achieved this by the deposition of an additional
multi-layer hBN flake on top of one of our graphene on hBN samples.
The resulting Raman spectrum is shown in Fig.~4c.
The spectrum displays three prominent peaks: the peak at 1367~cm$^{-1}$
is the E$_{2g2}$-mode of hBN. The G-line of graphene at 1583.7~cm$^{-1}$
is approximately in the same position as the G-line of graphene with
hBN on one side. This evidences that only a few additional charge impurities are
added during the deposition of the top-hBN flake. In contrast to the G-line,
the 2D-line of surrounded graphene around 2687.4~cm$^{-1}$ is considerably
blue-shifted by 7~cm$^{-1}$ compared to the 2D-line of one-sided 
graphene on hBN as shown in Fig.~4d. With respect to suspended graphene (2673.5~cm$^{-1}$)~\cite{berciaud} the blue-shift of the 2D-line of graphene on hBN roughly
doubles when a second hBN-layer is added on top. This is another indication
that the blue-shift has its origin in the screening dependence of the HOB
between K and M.

\section{Conclusion}
Our Raman measurements confirm 
 that hBN is a high-quality insulating substrate for graphene
with strongly reduced impurity charging. This is evidenced by the red-shift
of the G peak with respect to graphene on a standard SiO$_2$ substrate.
In contrast to the G-line, the 2D-line of graphene on hBN is blue-shifted.
We have shown that this change in the frequency of the highest-optical
branch is not a consequence of a direct (mechanical)
interaction between graphene and its substrate. It is rather an indirect,
electronic, effect mediated by the influence of the dielectric screening on the electronic structure of graphene. Usually, in three-dimensional metallic systems, 
the phonon-frequencies close to a Kohn anomaly depend only on the 
{\it internal} screening of the (bulk) material. Layered graphene is an 
example where (close to the Kohn anomaly at K), the frequencies of the 
highest optical branch depend on the {\it external} screening by the 
dielectric environment. This constitutes a new physical paradigm that
is worthwhile to be investigated also in other layered materials.
The electron-phonon coupling between the 
$\pi$-bands and the highest optical branch around $K$ can also impose a
limitation on the conductivity of graphene in the high current-limit
\cite{lazzeri2006}. Therefore, the dielectric screening of electron-phonon coupling
in 2-dimensional layered materials can play a general role for transport in
layered materials.

\section*{Acknowledgments}
Financial support  by the DFG (SPP-1459 and FOR-912) and the ERC is gratefully acknowledged.
A.M.-S. and L.W. acknowledge funding by the ANR (French National Research Agency) through project ANR-09-BLAN-0421-01.
Calculations were done at the IDRIS
supercomputing center, Orsay (Proj. No. 091827), and at the Tirant Supercomputer of the University of Valencia (group vlc44).

\section*{Appendix: Details on the calculations}

The DFT calculations are performed with the
code {\tt ABINIT}. Wave-functions are expanded in plane-waves with
an energy cutoff at 35 Ha. Core electrons are replaced by Trouiller-Martins
pseudopotentials. The periodic supercell comprises two layers of hBN and one 
layer of graphite with 20 a. u. of vacuum distance towards the neighboring
slab in order to keep the interaction between neighboring slabs small.
We use the experimental in-plane lattice constant of graphite, $a=2.46$~\AA,
and ``squeeze'' the hBN layers to the same value in order to keep the
calculations simple. 

The inter-layer spacing between adjacent hBN-layers is 3.33 {\AA} (experimental
value of bulk hBN) and between graphene and hBN 2.46 {\AA} (theoretical value,
obtained from geometry optimization of graphene on hBN). For the calculations
of ``isolated'' graphene, we use the same supercell, just removing the 
hBN-layers. The spacing between the graphene layers is thus the same in both
cases. In one case, there is vacuum between the layers, in the other case,
the vacuum is ``filled'' with hBN-layers.
The Brillouin zone is sampled with a $21 \times 21 \times 1$ 
$k$-point grid. The GW calculations have been done with the code 
{\tt Yambo}~\cite{yambo}, using the plasmon-pole
approximation for the dielectric constant. The convergence parameters
are the same as in Ref.~31. 
is used in the present work.
 The values for the electron-phonon coupling for pure graphene in Table~\ref{output} are
slightly different than in Ref.~31 
 due to the
different spacing between the layers which leads to a difference in the
(average) dielectric constant. 

The B3LYP exchange-correlation energy has the following form
\begin{eqnarray}
 E_{xc} & = & (1-a)(E_x^{LDA} + bE_x^{BECKE}) +aE_x^{HF} \nonumber \\
 & & + (1-c)E_c^{VWN} + cE_c^{LYP}, \nonumber
\end{eqnarray}
where $E_x^{BECKE}$ is the GGA exchange potential by Becke~\cite{becke},
$E_x^{HF}$ is the Hartree-Fock exchange potential,
$E_c^{VWN}$ is the LDA correlation potential by Vosko, Wilk, and 
Nusair~\cite{vosko}, and $E_c^{LYP}$ is the GGA correlation potential 
by Lee, Yang, and Parr~\cite{leeyangparr}.
The usual choice of the mixing parameters is $a=0.2, b=0.9, c=0.81$.
The parameter $a$ determines the admixture of Hartree-Fock exchange.
Increasing $a$ 
leads to a stiffening of the bonds (increasing thereby all optical phonon
frequencies). Increasing the parameter $c$ which 
determines the admixture of non-local correlation increases the
bond-strength as well. We therefore change the two parameters in opposite 
directions in order to keep the bond-strength constant.
Changing the parameter $a$ and $c$ then mimics, roughly, a change of 
screening.
Using the code {\tt CRYSTAL}~\cite{crystal}, we fit the parameters $a$
and $c$ such that we obtain the same values for the e-ph coupling matrix
elements, as calculated on the level of the GW-approximation.
For the case of isolated graphene, we obtain $a=0.157, c=0.1$,
for the case of surrounded graphene, we obtain $a=0.139, c=0.81$.
We emphasize, that this calculation is only meant to provide a
rough quantitative estimate and cannot reproduce the full physics
of the dielectric screening of the Kohn anomaly of the HOB.


\begin{thebibliography}{99}
\bibitem{latconst}
The measured (in-plane) lattice constants are $a$ = 2.4589(5) {\AA} for graphite~\cite{bas55} and $a$ = 2.504(2)~{\AA} for hexagonal boron nitride~\cite{sol95}.
\bibitem{random_note}
Note, that in the experiment, the carbon flake is deposited on 
top of hBN in random orientation. In this case, the band-gap remains 
closed~\cite{sachs} and the change in the Kohn anomaly of the HOB will be 
even less pronounced.
\bibitem{note_gap_opening}
Our GW-calculations are in line with the recent 
finding that the quasi-particle band-gap of ``graphone'' 
(partially hydrogenated graphene) is reduced by a hBN (or other dielectric) 
substrate~\cite{Kharche} and also with recent calculations of the band-gap
of MoS$_2$ in the presence of dielectric layers \cite{komsa}.
\bibitem{note_GWbands}
Due to the dielectric screening, the electron-electron (e-e) interaction in 
``sandwiched'' graphene is slightly reduced and thus the 
Fermi-velocity slightly decreases.
According to the double-resonance model, a reduction of the slope of the
$\pi$ bands leads to a shift of the phonon-wave vector away from K towards M
and thus to a blue-shift of the Raman 2D line. However, if one takes 
into account the effects of e-e interaction for the quasi-particle
band structure, one should also take into account excitonic effects
during the optical excitation of electron-hole (e-h) pairs. The increased
dielectric screening, which reduces the e-e interaction, also 
reduces the e-h interaction. Thus the two effects tend to cancel each other
and we assume that the Fermi-velocity reduction has a negligible
influence on the Raman 2D-line position.
\end{thebibliography}
\end{document}